\documentclass[aps,prd,preprintnumbers,nofootinbib,floatfix,11pt]{revtex4}

\usepackage{amsfonts,amsmath,amssymb,mathrsfs,graphicx,epsfig,color,amsthm} 
\usepackage{hyperref}

\allowdisplaybreaks[1]
\newcommand{\be}{\begin{equation}}
\newcommand{\ee}{\end{equation}}
\newcommand{\ba}{\begin{eqnarray}}
\newcommand{\ea}{\end{eqnarray}}

\theoremstyle{definition}

\usepackage{}

\usepackage[normalem]{ulem}

\begin{document}

\title{Maximum size of black holes in our accelerating Universe}
\author{Tetsuya Shiromizu$^{1,2}$, Keisuke Izumi$^{2,1}$, Kangjae Lee$^1$ and Diego Soligon$^1$} 

\affiliation{$^{1}$Department of Mathematics, Nagoya University, Nagoya 464-8602, Japan}
\affiliation{$^{2}$Kobayashi-Maskawa Institute, Nagoya University, Nagoya 464-8602, Japan} 

\begin{abstract}
\begin{center}
{\bf Abstract}
\end{center}
\noindent
In accordance with current models of the accelerating Universe as a spacetime with a positive cosmological constant, 
new results about a cosmological upper bound for the area of stable marginally outer trapped surfaces are found taking 
into account angular momentum, gravitational waves and matter. Compared to previous results which take into account only 
some of the aforementioned variables, the bound is found to be tighter, giving a concrete limit to the size of black holes 
especially relevant in the early Universe. 
\end{abstract}

\maketitle


\section{Introduction}

According to observations of our Universe, there are two accelerating expansion phases, 
the current phase and the very early stage. Such a Universe is approximately described as a spacetime 
with a positive cosmological constant. Considering the Einstein equation with a 
positive cosmological constant, the simplest solution is the de Sitter spacetime. 
Therefore, we can obtain useful insight into our Universe
by examining the characteristics of asymptotically de Sitter spacetimes. 
One interesting result is the presence of an upper bound for the area of 
event horizons. This is consistent with the intuitive feature that black holes should be smaller than 
the cosmological horizon, as we can see for the Schwarzschild-de Sitter solution. 
There are several works showing the existence of a cosmological upper bound for the area of an apparent horizon 
or so, in general spacetimes/axisymmetric spacetimes 
\cite{Shiromizu1993, Hayward1994b, Maeda1997, Gibbons1999, Simon2012, Simon2015, Khuri2017, Galloway2018}
(See section 4.4 in Ref.~\cite{Dain2018} for a review). 
In some of these the contribution from charge and angular momentum was also considered. 

In this paper, taking into account the contribution from every relevant variable we can think of, with the exception of charge, we re-examine 
the cosmological bound for the area of stable marginally outer trapped surface (stable MOTS) \cite{Simon2008} 
in realistic spacetimes with a positive cosmological constant. 
The analysis is quasi-local, that is, the asymptotic behavior of spacetime is not required to be imposed.
We can justify the omission of contribution from charge by the fact 
that it is unlikely that charged black holes exist in our neutral Universe 
\cite{Press1975, Ruffini2010}, so this simplifies our setup. 

This paper is organized as follows. In Sec.~II we present the setup and some basic concepts. 
In Sec.~III we derive the cosmological upper bounds for the sizes of black holes in spacetimes with a 
cosmological constant. Finally, we give a summary and discussion. The derivation of the key equation of the main result is presented separately in the appendix. 

\section{Setup and basics}
\label{basics}

Let us consider a compact 2-surface ($S$, $h_{ab}$) in a three-dimensional 
spacelike hypersurface ($\Sigma$, $q_{ab}$) in a four-dimensional 
spacetime ($M$, $g_{ab}$). Denoting the future-directed unit normal vector to $\Sigma$ in $M$ by $n^a$
and the outward-directed spacelike unit normal vector to $S$ in $\Sigma$ by $r^a$, the metric $g_{ab}$ is 
decomposed as 
\begin{equation}
g_{ab}=h_{ab}+r_ar_b-n_an_b=q_{ab}-n_an_b,
\end{equation}
and one can define the extrinsic curvatures of $\Sigma$ in $M$ and $S$ in $\Sigma$ as 
\begin{equation}
K_{ab}=q_a^cq_b^d \nabla_c n_d
\end{equation}
and 
\begin{equation}
k_{ab}=h_a^ch_b^d \nabla_c r_d,
\end{equation}
respectively, where $\nabla_a$ is the covariant derivative with respect to $g_{ab}$. We decompose $K_{ab}$ as 
\begin{equation}
K_{ab}=K_{(r)}r_ar_b+\kappa_{ab}+v_ar_b+v_br_a,
\end{equation}
where 
\begin{equation}
K_{(r)}:=K_{ab}r^ar^b,
\end{equation}
\begin{equation}
\kappa_{ab}:=h_a^ch_b^dK_{cd}
\end{equation}
and
\begin{equation}
v_a:=h_a^cr^bK_{bc}.
\end{equation}

After some manipulation, we obtain the derivative of $\theta_+=\kappa+k$ along the $r^a$-direction \cite{Andersson2010, Eichmair2012} (See Appendix for the full derivation) 
\begin{equation}
r^a \nabla_a \theta_+=-\frac{1}{2}\theta_{+ab}\theta^{ab}_+-\frac{1}{2}\theta_+^2+\theta_+ (K_{(r)}+\kappa)+\frac{1}{2}{}^{(2)}R+{\cal D}_a V^a
-V_aV^a-G_{ab}k^an^b, \label{der-theta}
\end{equation}
where $G_{ab}$ is the four-dimensional Einstein tensor, 
\begin{equation}
\theta_{+ab}:=\kappa_{ab}+k_{ab},
\end{equation}
\begin{equation}
V_a:=v_a-{\cal D}_a \ln \varphi,
\end{equation}
\begin{equation}
k^a:=n^a+r^a
\end{equation}
and ${\cal D}_a$ is the covariant derivative with respect to $h_{ab}$. 
$\varphi$ is the lapse function for the radial direction so that $q_b^c r^a \nabla_a r^b=-{\cal D}^c \ln \varphi$. 
Note that $k^a$ is a future-directed null vector and $V^a$ is spacelike vector tangent to $S$. 
Since $\theta_+=h^{ab}\nabla_a k_b=h^{ab}\theta_{+ab}$ holds, $\theta_+$ is the outgoing null expansion rate 
associated with the null vector $k^a$. 

\section{Cosmological upper bound for area of stable MOTS}

Following Ref.~\cite{Simon2008}, the stable marginally outer trapped surface (stable MOTS) is defined as a compact 2-surface $S$ satisfying 
\footnote{In Refs.~\cite{Hayward1994a, Hayward1994b} the concept of future trapping horizon was proposed instead, with the direction of the derivative in the second condition taken to be past-directed outgoing null.}
\begin{equation}
\theta_+ |_S = 0 \qquad {\rm and} \qquad r^a \nabla_a \theta_+ |_S \geq 0. \label{defTH}
\end{equation}
Then, taking the integral of Eq.~(\ref{der-theta}) over $S$ and 
using the Einstein equation $G_{ab}=8\pi T_{ab}-\Lambda g_{ab}$ with a positive cosmological constant $\Lambda$, 
we have  
\begin{equation}
\frac{1}{2}\int_S {}^{(2)}R \,dA \geq \Lambda A+\int_S \left[ V_a V^a+8\pi (\rho_++\rho_{+{\rm gw}})   \right]dA,
\label{int-ineq0}
\end{equation}
where $A$ is the area of the stable MOTS $S$, $\rho_+:=T_{ab}k^an^b=T_{ab}n^an^b+T_{ab}r^an^b$ and 
$\rho_{+{\rm gw}}$ is defined by 
\begin{equation}
8\pi \rho_{+{\rm gw}}:=\frac{1}{2}\tilde \theta_{+ab} \tilde \theta^{ab}_+=\frac{1}{2}( \tilde \kappa_{ab} \tilde \kappa^{ab}+\tilde k_{ab} \tilde k^{ab})
+\tilde \kappa_{ab} \tilde k^{ab}. \label{gw}
\end{equation}
The tilde stands for the traceless part, i.~e.~$\tilde \theta_{+ab}=\theta_{+ab}-(1/2)\theta_+ h_{ab}$.  
Note that $\rho_{+{\rm gw}}$ is positive definite. The last expression in the right-hand side of Eq.~(\ref{gw}) can be seen as 
a summation of the energy density and flux of gravitational waves as well as 
in $\rho_+$.  
Now, assuming that the dominant energy condition holds, that is $\rho_+ \geq 0$, one can see that the right-hand side of Eq.~(\ref{int-ineq0}) 
is non-negative.  From the Gauss-Bonnet theorem, this shows us that the topology of a stable MOTS is a 2-sphere 
\footnote{
Even for cases with a non-positive cosmological constant, Eq. \eqref{int-ineq0} holds. 
Then, if the integration in the right-hand side of Eq. \eqref{int-ineq0} is large enough that 
the right-hand side is positive, 
the topology of a stable MOTS is also shown to be a 2-sphere.
}, 
that is, 
$\int_S{}^{(2)}RdA=8\pi$. Thus, Eq.~(\ref{int-ineq0}) gives us 
\begin{equation}
A \leq \frac{4\pi}{\Lambda}-\frac{1}{\Lambda} \int_S \left[ V_a V^a+8\pi (\rho_++\rho_{+{\rm gw}})  \right]dA.
\label{int-ineq1}
\end{equation}
This can be regarded as a generalization of the result obtained by Gibbons for stable minimal surfaces \cite{Gibbons1999}. A minor arrangement 
gives us 
\begin{equation}
\int_S \left[ V^aV_a+8\pi \left(\rho_++\rho_{+{\rm gw}}  \right) \right] dA \leq 4\pi -\Lambda A. \label{VV-ineq}
\end{equation}

Bearing the cosmological setup in mind, one may define the surface-averaged total density $\bar \rho_{+{\rm tot}}$ as 
\begin{equation}
\bar \rho_{+{\rm tot}}=\frac{1}{A} \int_S \rho_{+{\rm tot}}dA,
\end{equation}
where $\rho_{+{\rm tot}}:=\rho_++\rho_{+{\rm gw}}$. Then, Eq.~(\ref{int-ineq1}) is simplified as 
\begin{equation}
A \leq \frac{4\pi}{\Lambda_+}-\frac{1}{\Lambda_+}\int_SV^aV_a dA,\label{int-ineq2}
\end{equation}
where 
\begin{equation}
\Lambda_+ := \Lambda+8\pi \bar \rho_{+{\rm tot}}.
\end{equation}
From the definition, $V^a$ includes the contribution from the angular momentum. Indeed, $v^a$ in $V^a$ is approximately expressed 
as $v^a=h^{ab}r^cK_{bc}=-h^{ab}n^c \nabla_b r_c \sim \partial_r g_{t \varphi}$, 
and non-trivial $g_{t \varphi}$ appears 
when spacetimes have a rotation, where $t$, $r$ and $\varphi$ are the time, radial and angular 
coordinates. Thus, one can see how the size of stable MOTS is limited by the presence of 
gravitational waves, matter, and angular momentum 
\footnote{
Even for cases with a non-positive cosmological constant, the result here is true if $\Lambda_+$ is positive.
}. 

As an application to the event horizon of a slowly rotating Kerr-de Sitter spacetime, 
we can evaluate the second term in the right-hand side of Eq.~(\ref{int-ineq2}). 
The metric of the Kerr-de Sitter spacetime is 
\begin{equation}
ds^2=-\frac{\zeta}{\rho^2}\left(dt-\frac{a \sin^2 \theta}{\lambda}d\phi \right)^2+\frac{\rho^2}{\zeta}dr^2+\frac{\rho^2}{\chi}d\theta^2
+\frac{\chi \sin^2 \theta}{\rho^2}\left(adt-\frac{r^2+a^2}{\lambda} d\phi \right)^2,
\end{equation}
where $\zeta=(r^2+a^2)(1-\Lambda r^2/3)-2mr$, $\rho^2=r^2+a^2\cos^2 \theta$, $\lambda=1+\Lambda a^2/3$ and 
$\chi=1+(\Lambda /3) a^2 \cos^2 \theta$. $a$ is the Kerr parameter and $m$ is the mass parameter. The unit normal vector $r^a$ to the horizon in the $t=$const.~slice is 
\begin{equation}
r_a=\frac{\rho}{{\sqrt \zeta}}(dr)_a
\end{equation}
and then one can read $\varphi$ as 
\begin{equation}
\varphi=\frac{\rho}{{\sqrt \zeta}}. 
\end{equation}
At the leading order, the components of $V_a$ are given by
\begin{equation}
V_\theta=-\partial_\theta \ln \varphi=\frac{a^2 \cos \theta \sin \theta}{\rho^2}=O(a^2)
\end{equation}
and
\begin{equation}
V_\phi=\frac{{\sqrt \zeta}}{\rho}K_{\phi r}=-\frac{3am}{r^2}\sin^2 \theta+O(a^3).
\end{equation} 
Note that the $\theta$-component of $V_a$ does not contribute to the evaluation at the leading order.  
Integrating over the event horizon, we have 
\begin{equation}
\int_{S_H}V_aV^adA=24\pi \frac{(am)^2}{r_H^4}+O(a^4).
\end{equation}
Notice that at the leading order there is no dependence on 
the cosmological constant. Although we can compute higher order terms of $a$, the expression would be rather 
complicated and it would be difficult to discern a physical or geometrical meaning. Nevertheless, since $(r_H^4/24)\int_{S_H}V_aV^adA$ 
gives us the angular momentum $(am)^2$ at the leading order, the above observation 
may prompt us to define the surface-averaged angular momentum as 
\footnote{In Ref.~\cite{Lee2022a}, $(8 \pi \bar J)^2 :=(A^2/6\pi) \int_{S}v^av_a dA$ was used. Note that $v^a$ is in the integrand, not $V^a$.}
\begin{equation}
(8 \pi \bar J)^2 :=\frac{A^2}{6\pi}\int_{S}V^aV_a dA. \label{ave-ang}
\end{equation}
Then, Eq.~(\ref{int-ineq2}) gives us 
\begin{equation}
A +\frac{6\pi (8\pi \bar J)^2}{\Lambda_+ A^2}  \leq \frac{4\pi}{\Lambda_+}. \label{int-ineq3}
\end{equation}
Due to the presence of the angular momentum, the area $A$ has a lower bound too. 
A minor rearrangement of Eq.~(\ref{int-ineq3}) gives us 
\begin{equation}
(8 \pi \bar J)^2 \leq \frac{2}{3} A^2 \left(1-\frac{\Lambda_+}{4\pi}A \right). \label{j2ineq-ours}
\end{equation}
This is similar to the inequality for the Komar angular momentum $J_K$ in 
axisymmetric spacetimes \cite{Simon2015}:
\begin{equation}
(8 \pi J_K)^2 \leq A^2 \left(1-\frac{\Lambda}{4\pi}A \right) 
\left(1- \frac{\Lambda}{12\pi}A \right). \label{j2ineq-simon}
\end{equation}
Note however that $\bar J$ does not coincide with the Komar angular momentum $J_K$, even for axisymmetric spacetimes. 
Indeed, one can show the following inequality \cite{Lee2022a, Lee2022b}
\begin{equation}
\bar J^2 \leq \left( \frac{{\cal R}_A}{{\cal R}_\phi}\right)^4J^2_K,
\end{equation}
where ${\cal R}_\phi$ is defined by \footnote{Considering the inverse mean curvature flow for 
the spacelike hypersurface $\Sigma$, one can show that ${\cal R}_\phi \geq {\cal R}_A$ for the oblate cases and 
${\cal R}_\phi \leq {\cal R}_A$ for the prolate cases \cite{Lee2022a}. Only for the spherically symmetric case 
${\cal R}_\phi ={\cal R}_A$ holds.}
\begin{equation}
{\cal R}_\phi^4:=\frac{3}{8\pi} \int_S \phi_a \phi^a dA
\end{equation}
with $\phi^a$ the axisymmetric Killing vector. We used the fact that 
$\int_S V^a \phi_a dA=\int_S v^a \phi_a dA $. 
For the Kerr-de Sitter spacetime, $J_K=am/\lambda$ (for example, see Ref.~\cite{Simon2015}). 
Then, one can see that $J_K = \bar J+O(a^2)$ for small $a$ cases.


\section{Summary and discussion}
\label{summary}

When one thinks of gravitational collapse of matter spreading over the cosmological horizon, 
a part of the matter of the outside region of the cosmological horizon does not collapse due to the cosmological expansion. 
Then, one can expect that the size of a black hole is smaller than that of the Universe, and we aim to show this. 
In particular, we derive a new inequality for the area of stable MOTS, such that the size of a black hole is limited by 
the presence of matter, gravitational waves and angular momentum. A main application of 
this consequence is 
to give certain considerations to the population and features of primordial black holes formed in the very early stage of our Universe. 
When one considers standard cosmology, the surface-averaged density (plus flux) $\bar \rho_+=(1/A)\int_ST_{ab}n^ak^bdA$ 
is composed of a homogeneous-isotropic part and cosmological perturbations, and the latter imply only tiny effects at the second order for the 
inequality. However, in non-linear events such as the black hole formation, the argument would be drastically changed. 
Note that even if $\Lambda=0$, we would still have a cosmological upper bound for stable MOTS due to the presence of a 
non-zero $\bar \rho_+$ in the expanding Universe, so that $\Lambda_+ \simeq 8\pi\bar \rho_+$ in standard cosmology as expected 
(See Ref.~\cite{Galloway2018} for a similar argument). 

One may be interested in higher dimensional cases. For the $n$-dimensional cases, Eq. (\ref{int-ineq0}) becomes 
\begin{equation}
\frac{1}{2}\int_S {}^{(n-1)}R \,dA \geq \Lambda A+\int_S \left[ V_a V^a+8\pi (\rho_++\rho_{+{\rm gw}})   \right]dA. 
\end{equation}
This inequality does hold regardless of dimensions and the sign of the cosmological constant. 
For $n=3$, we can evaluate the left-hand side from the Gauss Bonnet theorem, that is, $\int_S {}^{(n-1)}R \,dA=8 \pi (1-g)$, 
where $g$ is the number of genus. However, for $n \ge 4$, the Gauss-Bonnet theorem does not work, and 
thus the value and sign of the left-hand side cannot be fixed. It gives us an obstacle to show the 
upper bound for the area of the stable MOTS in higher dimensions. 

There are a few remaining issues. 
Equations (\ref{int-ineq1}) and (\ref{VV-ineq}) do not depend on 
the area-averaged quantities, which we introduce as a simplification. In reality, the exact physical meaning of the surface integral of $V^aV_a$ is 
unclear, although it can be definitely said that it includes the contribution from the angular momentum or a three-dimensional vector-type quantity. 
Furthermore, it is interesting to look at its variance for the evolution sequence of stable MOTSs. 
In the final phase of black hole formation, spacetime approaches a vacuum and stationary state near 
the black hole, and then the stable MOTS coincides with the cross section of the event horizon. This means that 
the second part of the integrand in Eq.~(\ref{int-ineq1}) decreases with evolution, but the first term does not 
and will be dominate in the integrand because of the existence of a possible angular momentum. In future studies, 
it would be nice to address these features. 

\acknowledgments

T.~S.~and K.~I.~are supported by Grant-Aid for Scientific Research from the Ministry of Education, 
Science, Sports and Culture of Japan (Nos. 17H01091, JP21H05182
JP21H05189). T.~S.~is also supported by JSPS Grants-in-Aid for Scientific Research (C) (JP21K03551). 
K.~I.~is also supported by JSPS Grants-in-Aid for Scientific Research (B) (JP20H01902)
and 
JSPS Bilateral Joint Research Projects (JSPS-DST collaboration) (JPJSBP120227705).

\appendix

\section{Sketch of derivation of Eq.~(\ref{der-theta})}

Here, we briefly describe the derivation of Eq.~(\ref{der-theta}). 

From the Codazzi equation, $q_a^b R_{bc}n^c=D_bK^b_a-D_aK$, we have 
\begin{eqnarray}
R_{ab}n^ar^b & = & r^aD_bK^b_a-r^aD_aK \nonumber \\
& = & -k_{ab}\kappa^{ab}+K_{(r)}k+{\cal D}_av^a+2v^a{\cal D}_a \ln \varphi-r^aD_a \kappa, \label{codazzi}
\end{eqnarray}
where $D_a$ is the covariant derivative with respect to $q_{ab}$. 
From the definition of the three dimensional Riemann tensor, ${}^{(3)}R_{ab}r^ar^b$ gives us 
\begin{eqnarray}
r^aD_ak & = & -k_{ab}k^{ab}-{}^{(3)}R_{ab}r^ar^b-\varphi^{-1}{\cal D}^2 \varphi \nonumber \\
& = & -\frac{1}{2}k_{ab}k^{ab}-\frac{1}{2}k^2+\frac{1}{2}{}^{(2)}R-\frac{1}{2}{}^{(3)}R-\varphi^{-1}{\cal D}^2 \varphi, \label{3rab}
\end{eqnarray}
where, in the second line, we used the double traced Gauss equation for $S$ in $\Sigma$. 

Using the double trace of the Gauss equation for $\Sigma$ in $M$, we have 
\begin{eqnarray}
G_{ab}n^an^b =  \frac{1}{2}\left({}^{(3)}R-K_{ab}K^{ab}+K^2 \right).\label{hami}
\end{eqnarray}
Using Eqs.~(\ref{codazzi}) and (\ref{3rab}), we can compute $r^aD_a (k+\kappa)$, and then, eliminating ${}^{(3)}R$ 
via Eq.~(\ref{hami}), we can derive Eq.~(\ref{der-theta}).

  
\end{document}